\documentclass[nofootinbib,prd]{revtex4}%
\usepackage{amsmath}
\usepackage{amsfonts}
\usepackage{amssymb}
\usepackage{graphicx}%
\begin{document}
\title{The cosmological constant as an eigenvalue of the Hamiltonian constraint in
Hor\v{a}va-Lifshits theory.}
\author{Remo Garattini}
\email{Remo.Garattini@unibg.it}
\affiliation{Universit\`{a} degli Studi di Bergamo, Facolt\`{a} di Ingegneria,}
\affiliation{Viale Marconi 5, 24044 Dalmine (Bergamo) Italy}
\affiliation{and I.N.F.N. - sezione di Milano, Milan, Italy.}

\begin{abstract}
In the framework of Hor\v{a}va-Lifshits theory, we study the eigenvalues
associated to the Wheeler-DeWitt equation representing the vacuum expectation
values associated to the cosmological constant. The explicit calculation is
performed with the help of a variational procedure with trial wave functionals
of the Gaussian type. We analyze both the case of the detailed balanced
condition and the case without it. In the case without detailed balance, we
find the existence of an eigenvalue depending on the coupling constants
$g_{2}$ and $g_{3}$ respectively and on the physical scale.

\end{abstract}
\maketitle

\section{Introduction}

The cosmological constant represents one of the challenges of this
century\cite{Lambda}. If one is tempted to compute it from the zero point
energy of some physical field, one discovers an enormous discrepancy with the
observation data. This discrepancy amounts to be of $10^{120}$ order of
magnitude: this is known as the cosmological constant problem. Many attempts
to explain such a discrepancy have been done, but they appear to be far from
being satisfactory. Recently, Ho\v{r}ava proposed a modification of Einstein
gravity motivated by the Lifshitz theory in solid state physics\cite{Horava}%
\cite{Lifshitz}. Such modification allows the theory to be power-counting
ultraviolet (UV)-renormalizable and should recover general relativity in the
infrared (IR) limit. Nevertheless Ho\v{r}ava-Lifshitz (HL) theory is
non-covariant. Indeed, in this approach space and time exhibit Lifshitz scale
invariance of the form%
\begin{equation}
t\rightarrow\ell^{z}t\ \text{\textrm{and}}\ x^{i}\rightarrow\ell x^{i}%
\end{equation}
with $z\geq1$. $z$ is called the dynamical critical exponent and in the
present case it is fixed to $z=3$. The breaking of the 4D diffeomorphism
invariance allows a different treatment of the kinetic and potential terms for
the metric: from one side the kinetic term is quadratic in time derivatives of
the metric, form the other side the potential has high-order space
derivatives. In particular the UV behavior is dominated by the square of the
Cotton tensor of the $3D$ geometry by means of a $k^{6}$ contribution to the
propagator leading to a renormalizable power-counting theory. The original
HL\ theory is based on two assumptions -- detailed balance and
projectability\cite{Horava1}\footnote{Different aspects of HL theory are
discussed in
Refs.\cite{LMP,KK,Brandeberger,WW,BRM,GC,CBENS,YFCENS,GLENS,CCGNAPPMS,NoPro,Minamitsuji,CCO,Myung,MK,Konoplya,LiPang,NoPro1,NoPro2}%
.}. The projectability condition is a weak version of the invariance with
respect to time reparametrization and therefore to the Wheeler-DeWitt (WDW)
equation\cite{DeWitt}. Motivated by these interesting features, we ask
ourselves if the problem of the cosmological constant in HL theory has better
chances to be solved than in ordinary Einstein gravity\footnote{See also
Ref.\cite{ACS}, for a different approach on the cosmological constant problem
in HL\ theory.}. Indeed, the modification of the gravitational field at short
distances could produce new contributions that allow what in ordinary Einstein
gravity is forbidden. In ordinary Einstein gravity, there exists a well
accepted model connected with the cosmological constant: this is the
Friedmann-Robertson-Walker metric, whose line element is%
\begin{equation}
ds^{2}=-N^{2}dt^{2}+a^{2}\left(  t\right)  d\Omega_{3}^{2}. \label{FRW}%
\end{equation}
$d\Omega_{3}^{2}$ is the usual line element on the three sphere and $N$ is the
lapse function. In this background, we have simply%
\begin{equation}
R_{ij}=\frac{2}{a^{2}\left(  t\right)  }\gamma_{ij}\qquad\mathrm{and}\qquad
R=\frac{6}{a^{2}\left(  t\right)  }.
\end{equation}
For simplicity we consider the case in which $k=1$. The generalization to
$k=0,-1$ is straightforward. The WDW equation for such a metric is%
\begin{equation}
H\Psi\left(  a\right)  =\left[  -\frac{\partial^{2}}{\partial a^{2}}%
+\frac{9\pi^{2}}{4G^{2}}\left(  a^{2}-\frac{\Lambda}{3}a^{4}\right)  \right]
\Psi\left(  a\right)  =0. \label{WDW_0}%
\end{equation}
It represents the quantum version of the invariance with respect to time
reparametrization. If we define the following reference length $a_{0}%
=\sqrt{3/\Lambda}$, the Eq.$\left(  \ref{WDW_0}\right)  $ assumes the familiar
form of a one-dimensional Schr\"{o}dinger equation for a particle moving in
the potential%
\begin{equation}
U\left(  a\right)  =\frac{9\pi^{2}a_{0}^{2}}{4G^{2}}\left[  \left(  \frac
{a}{a_{0}}\right)  ^{2}-\left(  \frac{a}{a_{0}}\right)  ^{4}\right]
\end{equation}
with total zero energy and without a factor ordering. Eq.$\left(
\ref{WDW_0}\right)  $ can be cast into the following form%
\begin{equation}
\frac{\int\mathcal{D}a\Psi^{\ast}\left(  a\right)  \left[  -\frac{\partial
^{2}}{\partial a^{2}}+\frac{9\pi^{2}}{4G^{2}}a^{2}\right]  \Psi\left(
a\right)  }{\int\mathcal{D}a\Psi^{\ast}\left(  a\right)  \left[  a^{4}\right]
\Psi\left(  a\right)  }=\frac{3\Lambda\pi^{2}}{4G^{2}}, \label{WDW_1}%
\end{equation}
which appears to be an expectation value. In particular, it can be interpreted
as an eigenvalue equation with a weight factor on the normalization. The
application of a variational procedure in ordinary gravity with a trial wave
functional of the form%
\begin{equation}
\Psi=\exp\left(  -\beta a^{2}\right)
\end{equation}
shows that there is no real solution of the parameter $\beta$ compatible with
the procedure. The purpose of this paper is to obtain an eigenvalue equation
for the cosmological constant like the one in Eq.$\left(  \ref{WDW_1}\right)
$ but in HL theory with the help of the WDW equation. Nevertheless, as pointed
out by Mukohyama\cite{Mukohyama}, there are four versions of the theory:
with/without the detailed balance condition; and with/without the
projectability condition. In this paper, we will consider the problem with and
without detailed balanced condition.

\section{HL Gravity with detailed balanced condition}

\label{p1}With the assumption of the detailed balanced condition, the action
for the Ho\v{r}ava theory assumes the form%
\begin{equation}
S=\int_{\Sigma\times I}dtd^{3}x\left(  \mathcal{L}_{K}-\mathcal{L}_{P}\right)
,
\end{equation}
where%
\begin{equation}
\mathcal{L}_{K}=N\sqrt{g}\frac{2}{\kappa^{2}}\left(  K^{ij}K_{ij}-\lambda
K^{2}\right)
\end{equation}
is the Lagrangian kinetic term. The extrinsic curvature is defined as%
\begin{equation}
K_{ij}=\frac{1}{2N}\left\{  -\dot{g}_{ij}+\nabla_{i}N_{j}+\nabla_{j}%
N_{i}\right\}
\end{equation}
with $K=K^{ij}g_{ij}$ is its trace and $N_{i}$ is the shift function. For the
FRW metric, there is no shift function. The Lagrangian potential term
$\mathcal{L}_{P}$ is
\begin{equation}
N\sqrt{g}\left[  \frac{\kappa^{2}}{2w^{4}}C^{ij}C_{ij}-\frac{\kappa^{2}\mu
}{2w^{2}}\varepsilon^{ijk}R_{il}\nabla_{j}R_{k}^{l}+\frac{\kappa^{2}\mu^{2}%
}{8}R^{ij}R_{ij}-\frac{\kappa^{2}\mu^{2}}{8\left(  1-3\lambda\right)  }\left(
\frac{1-4\lambda}{4}R^{2}+\Lambda_{W}R-3\Lambda_{W}^{2}\right)  \right]  ,
\label{Lp}%
\end{equation}
where%
\begin{equation}
C^{ij}=\epsilon^{ikl}\nabla_{k}\left(  R_{l}^{j}-\frac{1}{4}R\delta_{l}%
^{j}\right)
\end{equation}
is the Cotton tensor. Plugging the FRW metric $\left(  \ref{FRW}\right)  $
into $\mathcal{L}_{P}$, we obtain%
\begin{equation}
\mathcal{L}_{P}=N\sqrt{g}\frac{\kappa^{2}\mu^{2}}{8}\left[  \frac{3}{\left(
1-3\lambda\right)  a^{4}}-\frac{6\Lambda_{W}}{\left(  1-3\lambda\right)
a^{2}}+\frac{3\Lambda_{W}^{2}}{\left(  1-3\lambda\right)  }\right]  \label{lp}%
\end{equation}
and the action for the potential part reduces to%
\begin{equation}
S_{P}=-\int_{\Sigma\times I}dtd^{3}x\mathcal{L}_{P}=-\frac{\kappa^{2}\mu^{2}%
}{8\left(  1-3\lambda\right)  }\int_{I}dtN2\pi^{2}a^{3}\left[  \frac{3}{a^{4}%
}-\frac{6\Lambda_{W}}{a^{2}}+3\Lambda_{W}^{2}\right]  .
\end{equation}
Concerning the kinetic part one gets%
\begin{equation}
\mathcal{L}_{K}=N\sqrt{g}\frac{2}{\kappa^{2}}\left(  K^{ij}K_{ij}-\lambda
K^{2}\right)  =a^{3}\sin^{2}\chi\sin\theta\frac{3}{N\kappa^{2}}\left(
\frac{\dot{a}}{a}\right)  ^{2}\left(  1-3\lambda\right)
\end{equation}
and the corresponding action is%
\begin{equation}
S_{K}=\int_{\Sigma\times I}dtd^{3}x\mathcal{L}_{K}=\int_{I}dt2\pi^{2}%
a^{3}\frac{3}{N\kappa^{2}}\left(  \frac{\dot{a}}{a}\right)  ^{2}\left(
1-3\lambda\right)  .
\end{equation}
Now, we can compute the canonical momentum and we find%
\begin{equation}
\pi_{a}=\frac{\delta S_{K}}{\delta\dot{a}}=\frac{12\pi^{2}}{\kappa^{2}}%
a\dot{a}\left(  1-3\lambda\right)  ,
\end{equation}
where we have set $N=1$. The resulting Hamiltonian is computed by means of the
usual Legendre transformation leading to%
\[
H=\int_{\Sigma}d^{3}x\mathcal{H}=\int_{\Sigma}d^{3}x\left[  \pi_{a}\dot
{a}-\mathcal{L}\right]
\]%
\begin{equation}
=\frac{\kappa^{2}\pi_{a}^{2}}{12\pi^{2}a\left(  1-3\lambda\right)  }%
+\frac{2\pi^{2}\kappa^{2}\mu^{2}}{8\left(  1-3\lambda\right)  }\left[
\frac{3}{a}-6\Lambda_{W}a+3\Lambda_{W}^{2}a^{3}\right]  \label{Ham}%
\end{equation}
and the classical constraint can be read off quite straightforwardly%
\begin{equation}
\pi_{a}^{2}+9\mu^{2}\pi^{4}\left[  1-2\Lambda_{W}a^{2}+\Lambda_{W}^{2}%
a^{4}\right]  =0. \label{clco}%
\end{equation}
Eq.$\left(  \ref{clco}\right)  $ has been analyzed in Ref.\cite{Minamitsuji}.
However, we are interested in the WDW equation associated to Eq.$\left(
\ref{clco}\right)  $, which is given by%
\begin{equation}
-\frac{\partial^{2}\Psi}{\partial a^{2}}+9\mu^{2}\pi^{4}\left[  1-2\Lambda
_{W}a^{2}+\Lambda_{W}^{2}a^{4}\right]  \Psi=0, \label{WDW}%
\end{equation}
where we have promoted the canonical momentum $\pi_{a}$ to an operator
$\pi_{a}=-i\partial_{a}$. Note that the classical constraint $\left(
\ref{clco}\right)  $ and the WDW equation $\left(  \ref{WDW}\right)  $ are
independent on the parameter $\lambda$ for the FRW metric. This means that the
limit of $\lambda\rightarrow1$, necessary to reproduce General Relativity,
does not constrain $\Lambda_{W}$ to be negative. It is useful to define the
following dimensionless variable $x=\sqrt{\Lambda_{W}}a$, then Eq.$\left(
\ref{WDW}\right)  $ becomes%
\begin{equation}
-\frac{\partial^{2}\Psi}{\partial x^{2}}\Lambda_{W}+9\mu^{2}\pi^{4}\left[
1-2x^{2}+x^{4}\right]  \Psi=0. \label{WDW0}%
\end{equation}
We compute the value of $\Lambda_{W}$, by adopting a variational technique
with a trial wave functional of the form%
\begin{equation}
\Psi=\exp\left(  -\beta x^{2}\right)  . \label{a}%
\end{equation}
To this purpose, Eq.$\left(  \ref{WDW0}\right)  $ must be transformed into an
expectation value computation%
\begin{equation}
\int_{0}^{\infty}dx\Psi^{\ast}\left\{  -\Lambda_{W}\frac{\partial^{2}%
}{\partial x^{2}}+9\mu^{2}\pi^{4}\left[  1-2x^{2}+x^{4}\right]  \right\}
\Psi=0, \label{WDW1}%
\end{equation}
with the eigenvalue $\Lambda_{W}$ implicitly defined by Eq.$\left(
\ref{WDW1}\right)  $. We find%
\begin{equation}
\int_{0}^{\infty}dx\Psi^{\ast}\Psi=\sqrt{\frac{\pi}{8\beta}}, \label{b}%
\end{equation}
representing the normalization of the wave function. The other terms become%
\begin{equation}
-\int_{0}^{\infty}dx\Psi^{\ast}\frac{\partial^{2}\Psi}{\partial x^{2}}%
=\int_{0}^{\infty}dx\Psi^{\ast}\left[  2\beta-4\beta^{2}x^{2}\right]
\Psi=2\beta\sqrt{\frac{\pi}{8\beta}}-4\beta^{2}\sqrt{\frac{\pi}{8}}\left(
\frac{1}{4}\beta^{-\frac{3}{2}}\right)  =\beta\sqrt{\frac{\pi}{8\beta}},
\label{c}%
\end{equation}%
\begin{equation}
-18\pi^{4}\mu^{4}\int_{0}^{\infty}dx\Psi^{\ast}x^{2}\Psi=-18\pi^{4}\mu
^{4}\sqrt{\frac{\pi}{8}}\left(  \frac{1}{4}\beta^{-\frac{3}{2}}\right)
\label{d}%
\end{equation}
and%
\begin{equation}
\int_{0}^{\infty}dx\Psi^{\ast}x^{4}\Psi=\sqrt{\frac{\pi}{8}}\left(  \frac
{3}{16}\beta^{-\frac{5}{2}}\right)  . \label{e}%
\end{equation}
Plugging Eqs.$\left(  \ref{b},\ldots\ref{e}\right)  $ into Eq.$\left(
\ref{WDW1}\right)  $, one gets%
\begin{equation}
\Lambda_{W}\left(  \beta\sqrt{\frac{\pi}{8\beta}}\right)  +9\pi^{4}\mu
^{2}\sqrt{\frac{\pi}{8\beta}}\left[  1-\frac{1}{2\beta}+\frac{3}{16\beta^{2}%
}\right]  =0.
\end{equation}
In the spirit of the variational procedure $\Lambda_{W}$ must be stationary
against arbitrary variation of the parameter $\beta$. Thus, we impose%
\begin{equation}
\frac{1}{9\pi^{4}\mu^{2}}\frac{d\Lambda_{W}\left(  \beta\right)  }{d\beta
}=0=\frac{1}{\beta^{4}}\left[  \beta^{2}-\beta+\frac{9}{16}\right]
\end{equation}
and we find that the solutions exist only for complex $\beta$. Things do not
change if we make an analytic continuation of parameter $\mu\rightarrow i\mu
$\cite{Minamitsuji,CCO,Myung,MK,Konoplya,LMP,KS}. Therefore, it appears that
in ordinary HL gravity we cannot find a real eigenvalue for the cosmological
parameter $\Lambda_{W}$. Things could change if we slightly modify the action
satisfying the detailed balanced condition\cite{Horava1} by adding an
appropriate IR term of the form%
\begin{equation}
\mathcal{L}_{P}^{small}=N\sqrt{g}\frac{\kappa^{2}\mu^{2}}{8\left(
1-3\lambda\right)  }\left[  \Lambda_{W}\alpha_{1}R+\alpha_{2}\Lambda_{W}%
^{2}\right]  .
\end{equation}
The addition of $\mathcal{L}_{P}^{small}$ affects Eq.$\left(  \ref{Ham}%
\right)  $ which becomes%
\begin{equation}
H=\frac{\kappa^{2}\pi_{a}^{2}}{12\pi^{2}a\left(  1-3\lambda\right)  }%
+\frac{\pi^{2}\kappa^{2}\mu^{2}}{4\left(  1-3\lambda\right)  }\left[  \frac
{3}{a}+6\Lambda_{W}\left(  1-\alpha_{1}\right)  a+\left(  3+\alpha_{2}\right)
\Lambda_{W}^{2}a^{3}\right]  .
\end{equation}
Simplifying somewhat, we obtain the new form of the WDW equation%
\begin{equation}
H\Psi=-\frac{\partial^{2}\Psi}{\partial x^{2}}\Lambda_{W}+3\pi^{4}\mu
^{2}\left[  3+6\left(  1-\alpha_{1}\right)  x^{2}+\left(  3+\alpha_{2}\right)
x^{4}\right]  \Psi=0, \label{WDW_a}%
\end{equation}
where we have used the dimensionless variable $x=\sqrt{\Lambda_{W}}a$. By
adopting the same methodology of Eq.$\left(  \ref{WDW1}\right)  $, one gets%
\begin{equation}
\Lambda_{W}\left(  \beta\sqrt{\frac{\pi}{8\beta}}\right)  +9\pi^{4}\mu
^{2}\sqrt{\frac{\pi}{8\beta}}\left[  1+\frac{1-\alpha_{1}}{2\beta}%
+\frac{\left(  3+\alpha_{2}\right)  }{16\beta^{2}}\right]  =0,
\label{LambdaDBC}%
\end{equation}
where we have used Eqs.$\left(  \ref{b},\ldots\ref{e}\right)  $. We now
compute the extreme of $\Lambda_{W}\equiv\Lambda_{W}\left(  \beta\right)  $ to
obtain%
\begin{equation}
\frac{1}{9\pi^{4}\mu^{2}}\frac{\partial\Lambda_{W}}{\partial\beta}=0=\frac
{1}{\beta^{4}}\left[  \beta^{2}+\beta\left(  1-\alpha_{1}\right)
+3\frac{\left(  3+\alpha_{2}\right)  }{16}\right]  ,
\end{equation}
whose solutions are$\allowbreak$%
\begin{equation}
\beta_{1,2}=\frac{\alpha_{1}-1}{2}\pm\frac{\sqrt{\Delta}}{2};\qquad
\Delta=\left(  1-\alpha_{1}\right)  ^{2}-3\frac{\left(  3+\alpha_{2}\right)
}{4}. \label{beta12}%
\end{equation}
In order to have normalizable solutions, we must impose the positivity of
$\beta_{1,2}$. Hence we have three cases: a) $\Delta>0$, b) $\Delta=0$ and c)
$\Delta<0$. The case c) involves complex solutions and therefore a wave
functional with an oscillating part. For this reason, it will be discarded. We
begin with the

\begin{description}
\item[Case $\Delta=0.$] We have two real coincident solutions determined by%
\begin{equation}
\beta_{1,2}=\frac{\alpha_{1}-1}{2}.
\end{equation}
However $\Delta=0$ means that%
\begin{equation}
1-\alpha_{1}=\pm\frac{1}{2}\sqrt{3\left(  3+\alpha_{2}\right)  }%
\end{equation}
which implies that%
\begin{equation}
\beta_{\pm}=\pm\frac{\sqrt{3\left(  3+\alpha_{2}\right)  }}{4}.
\end{equation}
Since only the positive root leads to a normalizable solution, we take
$\beta_{+}$ with the further condition of $\alpha_{2}>-3$, otherwise we have a
solution not normalizable. Plugging $\beta_{+}$ into Eq.$\left(
\ref{LambdaDBC}\right)  $, one gets%
\begin{equation}
\frac{\Lambda_{W}}{9\pi^{4}\mu^{2}}=-\frac{1}{3\beta_{+}}%
\end{equation}
and therefore will be discarded when $\lambda<1/3$ but it will be accepted
when $\lambda>1/3$. This is particular important when we approach the General
Relativity limit where $\lambda=1$ and when we are forced to have a negative
$\Lambda_{W}$.

\item[Case $\Delta>0.$] Here we have two real distinct solutions for the
parameter $\beta$ expressed by $\left(  \ref{beta12}\right)  $.We find the
following sub-cases:

\item[Sub-case $\alpha_{1}=1,$ $\alpha_{2}>-3.$] For this choice of the
parameters we find that $\beta_{2}<0$ and
\begin{equation}
\beta_{1}=\frac{\sqrt{3\left(  3+\alpha_{2}\right)  }}{2}\qquad\alpha_{2}>-3.
\end{equation}
Plugging $\beta_{1}$ into Eq.$\left(  \ref{LambdaDBC}\right)  $, we obtain%
\begin{equation}
\Lambda_{W}=-\frac{9\pi^{4}\mu^{2}}{\beta_{1}}\left[  1+\frac{\left(
3+\alpha_{2}\right)  }{16\beta_{1}^{2}}\right]
\end{equation}
and since its value is negative, this solution will be discarded when
$\lambda<1/3$ but it will be accepted when $\lambda>1/3$.

\item[Sub-case $\alpha_{1}\neq1,$ $\alpha_{2}=-3.$] We find that%
\begin{equation}
\beta_{1}=0\qquad\mathrm{and\qquad}\beta_{2}=\alpha_{1}-1
\end{equation}
with the further condition of having $\alpha_{1}>1$. $\beta_{1}$ must be
discarded because it leads to a non normalizable solution. Plugging $\beta
_{2}$ into Eq.$\left(  \ref{LambdaDBC}\right)  $, we obtain%
\begin{equation}
\Lambda_{W}=-\frac{9\pi^{4}\mu^{2}}{2\left(  \alpha_{1}-1\right)  },
\end{equation}
which is negative and therefore it will be discarded when $\lambda<1/3$ but it
will be accepted when $\lambda>1/3$.
\end{description}

In the next section, we abandon the detailed balanced condition to see if we
can obtain other different solutions.

\section{HL Gravity without detailed balanced condition}

In Refs.\cite{SVW,WM}, a more general potential form which avoids the detailed
balanced condition has been proposed. Its expression is%
\[
\mathcal{\tilde{L}}_{P}=N\sqrt{g}\left[  g_{0}\zeta^{6}+g_{1}\zeta^{4}%
R+g_{2}\zeta^{2}R^{2}+g_{3}\zeta^{2}R^{ij}R_{ij}+g_{4}R^{3}\right.
\]%
\begin{equation}
\left.  +g_{5}R\left(  R^{ij}R_{ij}\right)  +g_{6}R_{j}^{i}R_{k}^{j}R_{i}%
^{k}+g_{7}R\nabla^{2}R+g_{8}\nabla_{i}R_{jk}\nabla^{i}R^{jk}\right]  .
\label{lpnodb}%
\end{equation}
Couplings $g_{a}\left(  a=0\ldots8\right)  $ are all dimensionless and powers
of $\zeta$ are necessary to maintain such a property of $g_{a}$. Plugging the
FRW background into $\mathcal{\tilde{L}}_{P}$, one gets%
\begin{equation}
\mathcal{\tilde{L}}_{P}=N\sqrt{g}\left[  g_{0}\zeta^{6}+g_{1}\zeta^{4}\frac
{6}{a^{2}\left(  t\right)  }+\frac{12\zeta^{2}}{a^{4}\left(  t\right)
}\left(  3g_{2}+g_{3}\right)  +\frac{24}{a^{6}\left(  t\right)  }\left(
9g_{4}+3g_{5}+g_{6}\right)  \right]
\end{equation}
The term $g_{0}\zeta^{6}$ plays the role of a cosmological constant. In order
to make contact with the ordinary Einstein-Hilbert action in $3+1$ dimensions,
we set without loss of generality $g_{0}\zeta^{6}=2\Lambda$ and $g_{1}=-1$. In
case one desires to study the negative cosmological constant, the
identification trivially will be $g_{0}\zeta^{6}=-2\Lambda$ . After having set
$N=1$, the Legendre transformation leads to%
\begin{equation}
\mathcal{H}=\pi_{a}\dot{a}-\mathcal{L}_{K}+\mathcal{L}_{P}%
\end{equation}
and the Hamiltonian becomes%
\begin{equation}
H=\int_{\Sigma}d^{3}x\mathcal{H}=-\frac{\kappa^{2}\pi_{a}^{2}}{12\pi
^{2}a\left(  3\lambda-1\right)  }+2\pi^{2}a^{3}\left(  t\right)  \left[
2\Lambda-\frac{6\zeta^{4}}{a^{2}\left(  t\right)  }+\frac{12\zeta^{2}b}%
{a^{4}\left(  t\right)  }+\frac{24c}{a^{6}\left(  t\right)  }\right]  ,
\end{equation}
where%
\begin{equation}
\left\{
\begin{array}
[c]{c}%
3g_{2}+g_{3}=b\\
9g_{4}+3g_{5}+g_{6}=c
\end{array}
\right.  .
\end{equation}
The WDW equation can be easily extracted to give%
\begin{equation}
\pi_{a}^{2}\Psi+\frac{\left(  3\lambda-1\right)  }{\kappa^{2}}24\pi^{4}%
a^{4}\left(  t\right)  \left[  -2\Lambda+\frac{6\zeta^{4}}{a^{2}\left(
t\right)  }-\frac{12\zeta^{2}b}{a^{4}\left(  t\right)  }-\frac{24c}%
{a^{6}\left(  t\right)  }\right]  \Psi=0
\end{equation}
and adopting the same procedure of Eq.$\left(  \ref{WDW_a}\right)  $, we can
write%
\begin{equation}
\int_{0}^{\infty}da\Psi^{\ast}\left\{  \pi_{a}^{2}\Psi+\frac{\left(
3\lambda-1\right)  }{\kappa^{2}}24\pi^{4}a^{4}\left(  t\right)  \left[
-2\Lambda+\frac{6\zeta^{4}}{a^{2}\left(  t\right)  }-\frac{12\zeta^{2}b}%
{a^{4}\left(  t\right)  }-\frac{24c}{a^{6}\left(  t\right)  }\right]
\right\}  \Psi=0.
\end{equation}
We can rearrange the previous expression to obtain
\begin{equation}
\frac{\int_{0}^{\infty}da\Psi^{\ast}\left\{  \pi_{a}^{2}\Psi+\frac{\left(
3\lambda-1\right)  }{\kappa^{2}}144\pi^{4}\left[  \zeta^{4}a^{2}\left(
t\right)  -2\zeta^{2}b-4ca^{-2}\left(  t\right)  \right]  \right\}  \Psi}%
{\int_{0}^{\infty}da\Psi^{\ast}a^{4}\left(  t\right)  \Psi}=\frac{\left(
3\lambda-1\right)  }{\kappa^{2}}48\pi^{4}\Lambda. \label{Eig}%
\end{equation}
It is now clear the role of $\Lambda$. As a trial wave functional, we adopt
the same form of Eq.$\left(  \ref{a}\right)  $ and using the results of
Eqs.$\left(  \ref{b},\ldots\ref{e}\right)  $ we can write%
\begin{equation}
\frac{\beta\sqrt{\frac{\pi}{8\beta}}+\frac{\left(  3\lambda-1\right)  }%
{\kappa^{2}}144\pi^{4}\left[  \zeta^{4}\sqrt{\frac{\pi}{8}}\left(  \frac{1}%
{4}\beta^{-\frac{3}{2}}\right)  -2\zeta^{2}b\sqrt{\frac{\pi}{8\beta}}%
+4c\sqrt{2\pi\beta}\right]  }{\sqrt{\frac{\pi}{8}}\left(  \frac{3}{16}%
\beta^{-\frac{5}{2}}\right)  }=\frac{\left(  3\lambda-1\right)  }{\kappa^{2}%
}48\pi^{4}\Lambda,
\end{equation}
where we have used the following relationship to compute%
\begin{equation}
-c\int_{0}^{\infty}da\Psi^{\ast}a^{-2}\Psi=-c\int_{0}^{\infty}\frac{dx}%
{2\sqrt{x^{3}}}\exp\left(  -2\beta x\right)  =-\frac{c}{2}\Gamma\left(
-\frac{1}{2}\right)  \sqrt{2\beta}=c\sqrt{2\pi\beta}. \label{iv}%
\end{equation}
By simplifying somewhat, one gets%
\begin{equation}
\left(  1+16\tilde{c}c\right)  \beta^{3}+\tilde{c}\left[  \frac{\zeta^{4}}%
{4}\beta-2\beta^{2}\zeta^{2}b\right]  =9\pi^{4}\frac{\left(  3\lambda
-1\right)  }{\kappa^{2}}\Lambda, \label{Lambda}%
\end{equation}
where we have defined%
\begin{equation}
\tilde{c}=\frac{\left(  3\lambda-1\right)  }{\kappa^{2}}144\pi^{4}.
\end{equation}
By applying the variational procedure, one obtains%
\begin{equation}
\frac{d\Lambda}{d\beta}=3\left(  1+16\tilde{c}c\right)  \beta^{2}-4\tilde
{c}b\zeta^{2}\beta+\frac{\zeta^{4}\tilde{c}}{4}=0. \label{LB}%
\end{equation}
We now discuss the set of solutions taking under consideration the different
cases. We begin with the special case where

\begin{description}
\item[$1+16\tilde{c}c=0,$ $b\neq0.$] The solution of Eq.$\left(
\ref{LB}\right)  $ is%
\begin{equation}
\beta_{0}=\frac{\zeta^{2}}{16b}.
\end{equation}
Plugging $\beta_{0}$ into Eq.$\left(  \ref{Lambda}\right)  $, one gets%
\begin{equation}
\frac{\zeta^{6}}{8b}=\Lambda=\frac{g_{0}\zeta^{6}}{2},
\end{equation}
which implies%
\begin{equation}
\frac{1}{4b}=g_{0}. \label{bg0}%
\end{equation}
Another special case is

\item[$1+16\tilde{c}c\neq0,$ $b=0.$] In this situation, we find%
\begin{equation}
\beta_{\pm}=\mp\frac{\zeta^{2}\sqrt{\tilde{c}}}{2\sqrt{3\left(  1+16\tilde
{c}c\right)  }}. \label{bpm}%
\end{equation}
$\beta_{+}$ must be rejected because it leads to solutions which are not
normalizable. Plugging $\beta_{-}$ into Eq.$\left(  \ref{Lambda}\right)  $,
one gets%
\begin{equation}
\frac{8\sqrt{\tilde{c}}\zeta^{6}}{3\sqrt{3\left(  1+16\tilde{c}c\right)  }%
}=\Lambda=\frac{g_{0}\zeta^{6}}{2},
\end{equation}
which implies%
\begin{equation}
\frac{16\sqrt{\tilde{c}}}{3\sqrt{3\left(  1+16\tilde{c}c\right)  }}=g_{0}.
\label{g0Large}%
\end{equation}
If $c=1$, we can approximate the previous result to obtain%
\begin{equation}
\frac{\sqrt{3}}{9}=g_{0}. \label{c=1}%
\end{equation}
Instead for the particular case where $c=0$, we find from Eq.$\left(
\ref{bpm}\right)  $ that%
\begin{equation}
\beta_{\pm}=\mp\frac{\zeta^{2}\sqrt{\tilde{c}}}{2\sqrt{3}}%
\end{equation}
and plugging $\beta_{-}$ into Eq.$\left(  \ref{Lambda}\right)  $, one gets%
\begin{equation}
\frac{8\sqrt{3\tilde{c}}\zeta^{6}}{9}=\Lambda=\frac{g_{0}\zeta^{6}}{2},
\end{equation}
which implies%
\begin{equation}
\frac{16\sqrt{3\tilde{c}}}{9}=g_{0}, \label{Smallc}%
\end{equation}
namely we remain with a dependence on $\lambda$.
\end{description}

When the parameters satisfy the condition $1+16\tilde{c}c\neq0$ and $b\neq0$,
the general solution is%
\begin{equation}
\beta_{1,2}=\zeta^{2}\frac{2\tilde{c}b\pm\sqrt{\tilde{\Delta}}}{3\left(
1+16\tilde{c}c\right)  },
\end{equation}
with%
\begin{equation}
\tilde{\Delta}=\left(  2\tilde{c}b\right)  ^{2}-3\left(  1+16\tilde
{c}c\right)  \frac{\tilde{c}}{4}.
\end{equation}
As in section \ref{p1}, we can distinguish three cases: a) $\tilde{\Delta}>0$,
b) $\tilde{\Delta}=0$ and c) $\tilde{\Delta}<0$. Case c) is discarded by
imposing that we have two real solutions. We begin with the case:

\begin{description}
\item[$\tilde{\Delta}=0.$] We have two real coincident solutions%
\begin{equation}
\beta_{1,2}=\beta=\frac{\zeta^{2}}{8b} \label{beta}%
\end{equation}
when $4\tilde{c}b=\pm\sqrt{3\left(  1+16\tilde{c}c\right)  \tilde{c}}$. For
the positive sign in front of the square root, $\lambda>1/3$, $b>0$ and
$c>-1/16\tilde{c}$. Plugging Eq.$\left(  \ref{beta}\right)  $ into Eq.$\left(
\ref{Lambda}\right)  $, we find the relationship%
\begin{equation}
\frac{\zeta^{6}}{6b}=\Lambda=\frac{g_{0}\zeta^{6}}{2}\qquad\Longrightarrow
\qquad\frac{1}{3b}=g_{0}. \label{LambdaNDBC}%
\end{equation}
On the other hand, when $\lambda<1/3$, then necessarily $b<0$,
$c>1/16\left\vert \tilde{c}\right\vert $ and%
\begin{equation}
\beta_{1,2}=\beta=\frac{\zeta^{2}}{8b}<0
\end{equation}
and in this case we have no normalizable solutions. However, when $4\tilde
{c}b=-\sqrt{3\left(  1+16\tilde{c}c\right)  \tilde{c}}$, we can have two
further sub-cases: $\lambda>1/3$ and $b<0$ , which will be rejected because
this leads to a non normalizable solution. The other sub-case is given by
$\lambda<1/3$ and $b>0$. Here we can get solutions provided $c>1/16\left\vert
\tilde{c}\right\vert $.

\item[$\tilde{\Delta}>0.$] We have two real distinct solutions%
\begin{equation}
\beta_{1}=\zeta^{2}\frac{2\tilde{c}b+\sqrt{\tilde{\Delta}}}{3\left(
1+16\tilde{c}c\right)  }\qquad\mathrm{and\qquad}\zeta^{2}\frac{2\tilde
{c}b-\sqrt{\tilde{\Delta}}}{3\left(  1+16\tilde{c}c\right)  }=\beta_{2}
\label{b1b2}%
\end{equation}
when $4\tilde{c}b>\sqrt{3\left(  1+16\tilde{c}c\right)  \tilde{c}}$. The
general form of $\beta_{1}$ and $\beta_{2}$ is not very illuminating. It is
more useful consider even in this case particular choices of the parameters.

\item[Sub-case $b\neq0,$ $c=0.$] With this combination of the constants we get%
\begin{equation}
\beta_{1}=\zeta^{2}\frac{2\tilde{c}b+\sqrt{\left(  2\tilde{c}b\right)
^{2}-3\tilde{c}/4}}{3}\qquad\mathrm{and\qquad}\zeta^{2}\frac{2\tilde{c}%
b-\sqrt{\left(  2\tilde{c}b\right)  ^{2}-3\tilde{c}/4}}{3}=\beta_{2}.
\end{equation}
If $1\gg b$ and $\lambda<1/3$, we can write%
\begin{equation}
\beta_{1}\simeq\zeta^{2}\frac{-4\left\vert \tilde{c}\right\vert b+\sqrt
{3\left\vert \tilde{c}\right\vert }}{6}\qquad\mathrm{and\qquad}\beta_{2}%
\simeq\zeta^{2}\frac{-4\left\vert \tilde{c}\right\vert b-\sqrt{3\left\vert
\tilde{c}\right\vert }}{6}.
\end{equation}
Since $\beta_{2}<0$, this solution will be discarded. Plugging $\beta_{1}$
into Eq.$\left(  \ref{Lambda}\right)  $ and taking linear terms in $b$, one
gets%
\begin{equation}
\frac{2}{3}\left(  \frac{2\sqrt{3\left\vert \tilde{c}\right\vert }}%
{3}-\left\vert \tilde{c}\right\vert b\right)  \zeta^{6}=\Lambda=\frac
{g_{0}\zeta^{6}}{2},
\end{equation}
which implies%
\begin{equation}
g_{0}=\frac{4}{3}\left(  \frac{2\sqrt{3\left\vert \tilde{c}\right\vert }}%
{3}-\left\vert \tilde{c}\right\vert b\right)  .
\end{equation}
On the other hand when $4\sqrt{\tilde{c}}b\gg\sqrt{3}$, $\lambda>1/3$
necessarily, that it means that $\beta_{1}$ and $\beta_{2}$ become%
\begin{equation}
\beta_{1}\simeq\zeta^{2}\frac{4\tilde{c}b}{3}\qquad\mathrm{and\qquad}\beta
_{2}\simeq0.
\end{equation}
Plugging $\beta_{1}$ into Eq.$\left(  \ref{Lambda}\right)  $, we find
that$\allowbreak$%
\begin{equation}
\frac{9\Lambda}{144\zeta^{6}}\simeq-\frac{32}{27}b^{3}\tilde{c}^{2}
\label{neg1}%
\end{equation}
which implies%
\begin{equation}
g_{0}=-\frac{1024}{27}b^{3}\tilde{c}^{2}.
\end{equation}

\item[Sub-case $c\ll0$.] When $c$ is large and negative and $\left\vert
c\right\vert \gg b$, $\tilde{\Delta}$ can be approximated with%
\begin{equation}
\tilde{\Delta}\simeq\left(  2\tilde{c}b\right)  ^{2}+12\tilde{c}^{2}\left\vert
c\right\vert .
\end{equation}
Then the solutions in $\left(  \ref{b1b2}\right)  $ become%
\begin{equation}
\beta_{1}\simeq-\zeta^{2}\left(  \frac{b}{24\left\vert c\right\vert }%
+\frac{\sqrt{3\left\vert c\right\vert }}{24\left\vert c\right\vert }\right)
\qquad\mathrm{and\qquad}\beta_{2}\simeq-\zeta^{2}\left(  \frac{b}{24\left\vert
c\right\vert }-\frac{\sqrt{3\left\vert c\right\vert }}{24\left\vert
c\right\vert }\right)  .
\end{equation}
$\beta_{1}$ is always negative and therefore will be discarded, while
$\beta_{2}$ can be further reduced to%
\begin{equation}
\beta_{2}\simeq\frac{\zeta^{2}}{24}\sqrt{\frac{3}{\left\vert c\right\vert }}.
\end{equation}
Plugging $\beta_{2}$ into Eq.$\left(  \ref{Lambda}\right)  $, we find%
\begin{equation}
\frac{1}{9}\sqrt{\frac{3}{\left\vert c\right\vert }}\zeta^{6}=\Lambda
=\frac{g_{0}\zeta^{6}}{2},
\end{equation}
which implies%
\begin{equation}
\frac{2}{9}\sqrt{\frac{3}{\left\vert c\right\vert }}=g_{0}.
\end{equation}

\item[Sub-case $b\gg c\gg1.$] In this approximation we find%
\begin{equation}
\tilde{\Delta}\simeq\left(  2\tilde{c}b\right)  ^{2}-12\tilde{c}^{2}c
\end{equation}
Then the solutions in $\left(  \ref{b1b2}\right)  $ become%
\begin{equation}
\beta_{1}\simeq\zeta^{2}\left(  \frac{b}{12c}-\frac{1}{64b}\right)
\simeq\frac{\zeta^{2}b}{12c}\qquad\mathrm{and\qquad}\beta_{2}\simeq\frac
{\zeta^{2}}{64b}.
\end{equation}
Both $\beta_{1}$ and $\beta_{2}$ can be accepted to estimate $\Lambda$.
Plugging $\beta_{1}$ into Eq.$\left(  \ref{Lambda}\right)  $, we find%
\begin{equation}
\left(  \frac{b}{3c}-\frac{2b^{3}}{27c^{2}}\right)  \zeta^{6}=\Lambda
=\frac{g_{0}\zeta^{6}}{2}, \label{neg2}%
\end{equation}
which implies%
\begin{equation}
\frac{2b}{3c}-\frac{4b^{3}}{27c^{2}}=g_{0}.
\end{equation}
Instead, Plugging $\beta_{2}$ into Eq.$\left(  \ref{Lambda}\right)  $, we find%
\begin{equation}
\frac{7\zeta^{6}}{128b}=\Lambda=\frac{g_{0}\zeta^{6}}{2},
\end{equation}
which implies%
\begin{equation}
\frac{7}{64b}=g_{0}.
\end{equation}

\end{description}

\section{Conclusions}

In this paper, we have considered the recent proposal of H\v{o}rava to compute
the cosmological constant in a modified non-covariant gravity theory at the
Lifshitz point $z=3$. We have used the WDW equation as a backbone of all
calculations with the cosmological constant regarded as an
eigenvalue\cite{Remo}. We have analyzed the situation with detailed balanced
condition and without the detailed balanced condition. A variational approach
with Gaussian wave functions has been used to do calculation in practice. The
background geometry is a Friedmann-Robertson-Walker metric with $k=1$.
Concerning the detailed balanced condition, we have found an approximate
eigenvalue not in the original configuration but with a slight modification in
the IR region. The main reason is related to the fact that the original
parameter $\Lambda_{W}$ cannot be considered as an eigenvalue in this
formulation. What we have found is that we have only results in agreement with
a big cosmological constant, therefore with Planck era estimations. This means
that the cosmological constant problem in this approach is not solved. It is
interesting to note that only the dimensionless ratio $\Lambda_{W}/\mu^{2}$
comes into play. As regards the case without detailed balanced condition, we
obtain four main cases described by Eqs.$\left(  \ref{bg0},\ref{LambdaNDBC}%
\right)  $ and Eqs.$\left(  \ref{g0Large},\ref{c=1}\right)  $:

\begin{description}
\item[a)] $g_{0}$ is large and determined by the set of coupling constants
$\left(  g_{2},g_{3}\right)  $. For example, it could be fine tuned to Planck
era values and therefore to the order $10^{120}$. This implies that
$b=6g_{2}+g_{3}\ll1$. This can be achieved if $g_{2},g_{3}\ll1$ or
$6g_{2}\simeq-g_{3}$. When $g_{2},g_{3}\ll1$ we fall into a perturbative
regime, while when $6g_{2}\simeq-g_{3}$ this could not be the case. In this
respect, this version of HL theory and the version with detailed balanced
condition behave in the same way, except eventually for the smallness of the
coupling constants $g_{2}$ and $g_{3}$.

\item[b)] $g_{0}$ is order of unity or less and determined by the set of
coupling constants $\left(  g_{2},g_{3}\right)  $. It can be fine tuned to the
values coming from observation. This implies that $b=6g_{2}+g_{3}\simeq1$,
that it means that the set $\left(  g_{2},g_{3}\right)  $ can be into the
perturbative region. For example, it is sufficient to take the couple $\left(
g_{2}=1/12,g_{3}=1/2\right)  $. Indeed since in case \textbf{a)}, we started
with $g_{0}\simeq10^{120}$, we do not need to obtain $g_{0}\simeq10^{-120}$ as
a final result.

\item[c)] $g_{0}$ is large and determined by the set of coupling constants
$\left(  g_{4},g_{5},g_{6}\right)  $ in Eq.$\left(  \ref{g0Large}\right)  $.
It can be fine tuned to be of the order of $10^{120}$. This can be realized
when the combination $c=9g_{4}+3g_{5}+g_{6}<0$ and even if $1+16\tilde{c}%
c\neq0$, nothing prevents to consider the case in which $1+16\tilde{c}c$ is
small. Since we are dealing with three coupling constants, it is not trivial
to do a discussion similar to the cases in \textbf{a)} and \textbf{b)},
because we have more combinations. However, when one of the constants vanish
one can repeat the same analysis of case \textbf{a)} and one discover three
other sub-sub-cases

\item[d)] $g_{0}$ is small and determined by the set of coupling constants
$\left(  g_{4},g_{5},g_{6}\right)  $ in Eq.$\left(  \ref{c=1}\right)  $. This
can be achieved if $c\rightarrow1$ then $g_{0}<1$. Of course, even in this
case, we do not need to obtain $g_{0}\simeq10^{-120}$ as a final result. The
same considerations of the case \textbf{c)} can be applied here. Nevertheless,
to have a better comparison with observation one should adopt the units in
which $\left[  dx\right]  =\left[  dt\right]  $ where the speed of light is
equal to one\cite{SVW,WM}. We can observe that a big difference between HL
theory with and without balanced condition is that the latter admits a
possible transition from big values to small values of the cosmological
constant. However, it is likely that an improvement could be obtained by
introducing a renormalization group equation for the various coupling constant
and eventually a running $\Lambda$. Another improvement for having bounds on
parameters can be extracted from the equation of state $p=\omega\rho$ with
$\omega=-1$. Using the Gaussian wave function's value of the exponent $\beta$
obtained by the variational procedure into the equation of state, we can
extract other informations on the coupling constants. Even if a further and
deeper investigation is needed, it is interesting to observe that the set of
coupling constants $\left(  g_{2},g_{3}\right)  $ and the set $\left(
g_{4},g_{5},g_{6}\right)  $ seem to be disconnected one each other when we
estimate $g_{0}$. Needless to say that the analysis has been done without any
matter field and also this addition could improve the final result.
\end{description}

\end{document}